\documentclass[journal,onecolumn]{IEEEtran}

\usepackage[utf8]{inputenc}
\usepackage{graphicx}
\usepackage{amsmath,amssymb}

\begin{document}

\title{Securing Open RAN: A Survey of Cryptographic Challenges and Emerging Solutions for 5G}

\author{\IEEEauthorblockN{Ryan Barker, Fatemeh Afghah} \\
    \IEEEauthorblockA{Holcombe Department of Electrical and Computer Engineering, \\
    Clemson University, Clemson, SC, USA \\
    Emails: \{rcbarke, fafghah\}@clemson.edu}
}

\maketitle

\begin{abstract}
The advent of Open Radio Access Networks (O-RAN) introduces modularity and flexibility into 5G deployments but also surfaces novel security challenges across disaggregated interfaces. This literature review synthesizes recent research across thirteen academic and industry sources, examining vulnerabilities such as cipher bidding-down attacks, partial encryption exposure on control/user planes, and performance trade-offs in securing O-RAN interfaces like E2 and O1. The paper surveys key cryptographic tools—SNOW-V, AES-256, and ZUC-256—evaluating their throughput, side-channel resilience, and adaptability to heterogeneous slices (eMBB, URLLC, mMTC). Emphasis is placed on emerging testbeds and AI-driven controllers that facilitate dynamic orchestration, anomaly detection, and secure configuration. We conclude by outlining future research directions, including hardware offloading, cross-layer cipher adaptation, and alignment with 3GPP TS 33.501 and O-RAN Alliance security mandates, all of which point toward the need for integrated, zero-trust architectures in 6G.
\end{abstract}


\section{Introduction}
\label{sec:intro}

Open Radio Access Network (O-RAN) initiatives promise to revolutionize cellular communications by disaggregating traditionally monolithic base station components into open, multi-vendor elements~\cite{Understanding_ORAN}. Through standardized interfaces (e.g., \texttt{E2}, \texttt{O1}), operators can flexibly integrate hardware and software from diverse vendors, reducing deployment costs and promoting rapid innovation. However, this openness also introduces critical security challenges, including novel attack surfaces and uncertain trust boundaries across heterogeneous components.

Recent prototype deployments have demonstrated the versatility of O-RAN in both commercial and experimental settings~\cite{oaic,x5g}. These platforms often incorporate advanced orchestration schemes, real-time controllers (e.g., near-RT RIC), and machine learning-based applications (xApps) for tasks like resource allocation and interference management~\cite{orchest_ran}. Despite clear benefits in adaptability and scalability, the increased exposure of internal control signals—once hidden inside proprietary hardware—amplifies the risk of malicious manipulation and data leakage. For instance, adversaries may exploit partially encrypted control-plane exchanges or force cipher downgrades to eavesdrop on user-plane traffic.

This review discusses how the latest research efforts address these O-RAN security gaps. Specifically, we focus on (i) the complexities of securing open interfaces under strict performance constraints, (ii) emerging cryptographic primitives (e.g., SNOW-V, ZUC-256) to meet 5G’s 256-bit security requirements, and (iii) cross-layer solutions coupling slice-aware policies with hardware offloading techniques. The synergy between these techniques and advanced artificial intelligence pipelines enables closed-loop network optimization while preserving confidentiality and integrity. Nonetheless, questions remain about cost-effective implementations, side-channel defenses, and global standardization efforts.

We proceed by examining the technical foundations of O-RAN’s architecture and interfaces, highlighting key points of vulnerability. Subsequently, we analyze existing literature on threat models, cryptographic solutions, and reported overheads or trade-offs. Finally, we outline future research directions, including zero-trust RAN designs and deeper integration of security orchestration with multi-domain slice management. 

\section{O-RAN Architecture and Interfaces}
\label{sec:oran_arch}

Open Radio Access Network (O-RAN) initiatives introduce a modular design that disaggregates radio resources and virtualized functions to enhance flexibility and promote multi-vendor compatibility. As depicted in Figure~1 (located at the end of this document), the transition from monolithic RAN solutions to O-RAN involves splitting traditional base station elements into an O-RU (Radio Unit), O-DU (Distributed Unit), and O-CU (Central Unit), as well as adding a near-real-time RIC (RAN Intelligent Controller) and a Service Management and Orchestration (SMO) framework~\cite{Understanding_ORAN,dApps,x5g}. This architectural evolution empowers operators to mix-and-match hardware and software from diverse sources, allowing tight resource control and more agile deployments.

The O-RU handles lower physical-layer functions and interacts directly with user equipment over the air interface. The O-DU, typically residing at an edge data center, manages higher-layer PHY and MAC operations, forwarding control-plane signaling to the O-CU for RRC/radio resource scheduling tasks. The near-RT RIC introduces a software-defined layer for real-time optimization and ML-driven control, while the SMO and non-RT RIC components coordinate longer-term resource orchestration. By standardizing these disaggregated blocks, O-RAN fosters interoperability and rapid innovation.

Projects such as OAIC~\cite{oaic} and X5G~\cite{x5g} demonstrate how open, end-to-end 5G O-RAN testbeds accelerate development and experimentation. These platforms include a combination of split deployments (e.g., \texttt{7.2x}, \texttt{F1}, or \texttt{E2}) and orchestrated control layers (e.g., near-RT RIC, SMO) exposed to external 5G core networks. By offering programmable interfaces, they facilitate rapid prototyping of custom scheduling, real-time xApps, and zero-touch service management.

A key O-RAN innovation is the introduction of the near-RT RIC and non-RT RIC~\cite{dApps,orchest_ran}. The near-RT RIC supports \emph{time-sensitive} optimization (on the order of 10\,ms to 1\,s) for tasks like dynamic radio resource allocation. Meanwhile, the non-RT RIC provides higher-level policy and long-term analytics. Together, they enable closed-loop control pipelines that incorporate advanced AI/ML algorithms (via xApps or dApps) for interference mitigation, load balancing, and adaptive slicing, crucially improving end-user Quality of Service while retaining open, vendor-neutral RAN components.

\section{Key Security Challenges in O-RAN}
\label{sec:security_challenges}

Despite its architectural flexibility and innovation potential, O-RAN introduces a variety of security risks due to its open interfaces, disaggregated control, and multi-vendor design. This section outlines several emerging threats observed in recent research.

Karakoc et al.~\cite{never_let_me_down_again} describe downgrade attacks in which adversaries manipulate encryption negotiation to force fallback to insecure cipher suites like NEA0 (null encryption). These “bidding-down” exploits can occur during attach or handover procedures in both 5G Standalone (SA) and Non-Standalone (NSA) configurations, enabling attackers to intercept user data or inject malicious signaling. Without robust enforcement of cipher selection policies, such attacks compromise both confidentiality and control-plane integrity.

Zhao et al.~\cite{messaging_insecurities} expose vulnerabilities in 5G messaging and Rich Communication Services (RCS), where weak or missing encryption of signaling elements allows for impersonation, session hijacking, and partial leakage of user-plane data. These risks are particularly pronounced when RCS elements are integrated with control-plane mechanisms in O-RAN architectures, increasing the attack surface for adversaries targeting messaging states.

Securing O-RAN's open interfaces, particularly E2 and O1, presents significant implementation and performance trade-offs. Groen et al.~\cite{securing_oran} evaluate encryption strategies and find substantial latency and processing overhead when IPsec or TLS is applied to every control-plane message. Furthermore, practical evaluations~\cite{oran_interface_security} indicate that encryption is inconsistently applied across vendor deployments, leaving critical links like E2 partially or fully exposed. These findings highlight the difficulty of achieving uniform, low-latency security in disaggregated RAN environments.

In addition to transport-layer risks, stream ciphers used in RAN encryption face side-channel threats. Recent work on SNOW-V~\cite{snow_sca,snow_v_fpga,snow_v} demonstrates that power and electromagnetic (EM) side-channel analysis can leak cryptographic key material in fewer than 50 traces. These attacks exploit implementation-level leakage in the cipher’s LFSR updates or AES-based FSMs, emphasizing the need for constant-time logic and masking in hardware and software implementations used within O-RAN components.

\section{Evolving Cryptographic Tools in 5G}
\label{sec:evolving_crypto}

As O-RAN embraces open interfaces and disaggregated components, robust encryption mechanisms become essential to maintain both performance and security. This section surveys three primary 256-bit ciphers adopted or proposed for 5G contexts, along with comparative benchmarks and known trade-offs.

Ekdahl et al.~\cite{snow_v} introduced SNOW-V as an evolution of the SNOW family, tailored for high-speed 5G applications. It uses a linear feedback shift register (LFSR) in tandem with AES-like round functions to produce a 128-bit keystream per iteration. Further studies have explored lightweight FPGA implementations of SNOW-V~\cite{snow_v_fpga}, targeting resource-constrained or virtualized environments that require fast, inline encryption. Concurrently, side-channel analysis on SNOW-V~\cite{snow_sca} reveals potential leakage from LFSR updates, underscoring the need for hardware masking and constant-time logic in real deployments.

AES-256 remains one of the most ubiquitous block ciphers for 5G and beyond, largely due to its standardized security and broad industry support. Wei et al.~\cite{hpc_riscv} implement AES-256 on RISC-V to compare performance with other stream ciphers, demonstrating that without hardware acceleration, AES-256 can exhibit higher latency than specialized stream ciphers. Nonetheless, its widespread adoption, existing hardware support (e.g., AES-NI on x86), and compatibility with advanced modes like AES-GCM make it a mainstay for confidentiality and authentication in open RAN deployments.

ZUC-256 is another 256-bit stream cipher, particularly recognized in certain regions and standardized for 5G scenarios. Like SNOW-V, it is designed to provide high throughput with a relatively compact footprint. In the RISC-V benchmark of Wei et al.~\cite{hpc_riscv}, ZUC-256 shows favorable performance under software-only conditions, although it lacks the parallel AES-based structure of SNOW-V. As 5G systems diversify across global markets, ZUC-256 remains an essential candidate for securing O-RAN interfaces, especially in jurisdictions mandating its usage.

Multiple studies highlight distinct performance and security considerations for these ciphers. Groen et al.~\cite{securing_oran} and the open interface analyses of~\cite{oran_interface_security} note that implementing full encryption on critical interfaces can introduce measurable overhead, while specialized designs for SNOW-V in FPGA~\cite{snow_v_fpga} reduce latencies significantly. Overall, Wei et al.~\cite{hpc_riscv} underscore that ZUC-256 and SNOW-V may outperform AES-256 in software-only RISC-V environments, yet each cipher faces vulnerabilities unless side-channel defenses and uniform interface-level encryption are consistently applied.

\section{Future Research Directions and Potential Approaches}
\label{sec:future_directions}

Building upon current efforts to secure O-RAN deployments, several future directions emerge that seek to balance performance, automation, and cryptographic strength.

One promising avenue is the integration of security orchestration within the AI-based RAN control framework, where dApps and xApps~\cite{dApps,oaic} deployed at the O-DU, O-CU, and near-real-time RIC monitor and respond to anomalies in cipher negotiation, signaling integrity, or partial encryption. These intelligent controllers could leverage real-time metrics to detect downgrades to insecure modes (e.g., NEA0) or unencrypted control-plane elements, triggering automated countermeasures or key renegotiation.

To enforce strong security without compromising latency or throughput, future systems may adopt inline hardware-based encryption, offloading cryptographic operations to dedicated FPGA or SoC cores~\cite{oran_interface_security,snow_v_fpga}. These accelerators, placed at critical nodes like the O-DU and O-CU, can process user and control-plane traffic at line speed. Designs must also be hardened against side-channel attacks~\cite{snow_sca}, integrating masking, constant-time logic, and physically isolated key storage to comply with zero-trust paradigms in O-RAN and 6G.

Dynamic security adaptation across network slices presents another key opportunity. By linking slice-level policies to cipher selection, RAN systems could optimize for both performance and security. For instance, URLLC flows with tight latency budgets might prefer SNOW-V for its low-overhead keystream generation~\cite{snow_v,hpc_riscv}, while eMBB slices can leverage AES-256-GCM for bulk data encryption. This adaptive approach must consider trade-offs across layers, ensuring that security policies align with application-level SLAs.

Finally, harmonizing encryption mandates across global standards bodies remains an open challenge. While 3GPP TS 33.501~\cite{Understanding_ORAN} and recent security analyses~\cite{never_let_me_down_again} underscore the importance of end-to-end 256-bit encryption, enforcement varies across vendors and geographies. As O-RAN Alliance specifications evolve, stronger guidance on interface-level protections and key management policies will be critical. Future research must explore how to unify these expectations into a coherent global framework that ensures robust security without fragmenting the ecosystem.

\section{Conclusion}
\label{sec:conclusion}

Open RAN offers a transformative opportunity for programmable, multi-vendor wireless systems, but its openness also introduces profound security challenges across disaggregated components and interfaces. To protect control and user-plane traffic, next-generation RAN systems increasingly rely on advanced cryptographic primitives such as SNOW-V, AES-256, and ZUC-256. Experimental platforms like OAIC and X5G~\cite{oaic,x5g}, along with intelligent orchestration tools like dApps and xApps~\cite{dApps,oaic}, demonstrate the growing role of AI in maintaining secure, real-time operation. Simultaneously, cryptographic innovations~\cite{snow_v,snow_v_fpga,snow_sca} and comparative benchmarking~\cite{hpc_riscv} expose performance limitations and the complexity of implementing robust side-channel defenses.

There is a growing consensus around the need for hardware-accelerated, slice-aware, 256-bit encryption tailored to the performance constraints of URLLC, eMBB, and mMTC slices. Future 6G networks will likely demand deeper integration among RAN disaggregation, zero-trust security models, and AI-driven orchestration. Achieving this synergy will require continued exploration of dynamic cipher adaptation, secure control-plane signaling, and global standards harmonization.

\ifCLASSOPTIONcaptionsoff
  \newpage
\fi

\nocite{*}
\bibliographystyle{IEEEtran}
\bibliography{References}

\begin{figure}[!h]
    \centering
    \includegraphics[width=\columnwidth, ]{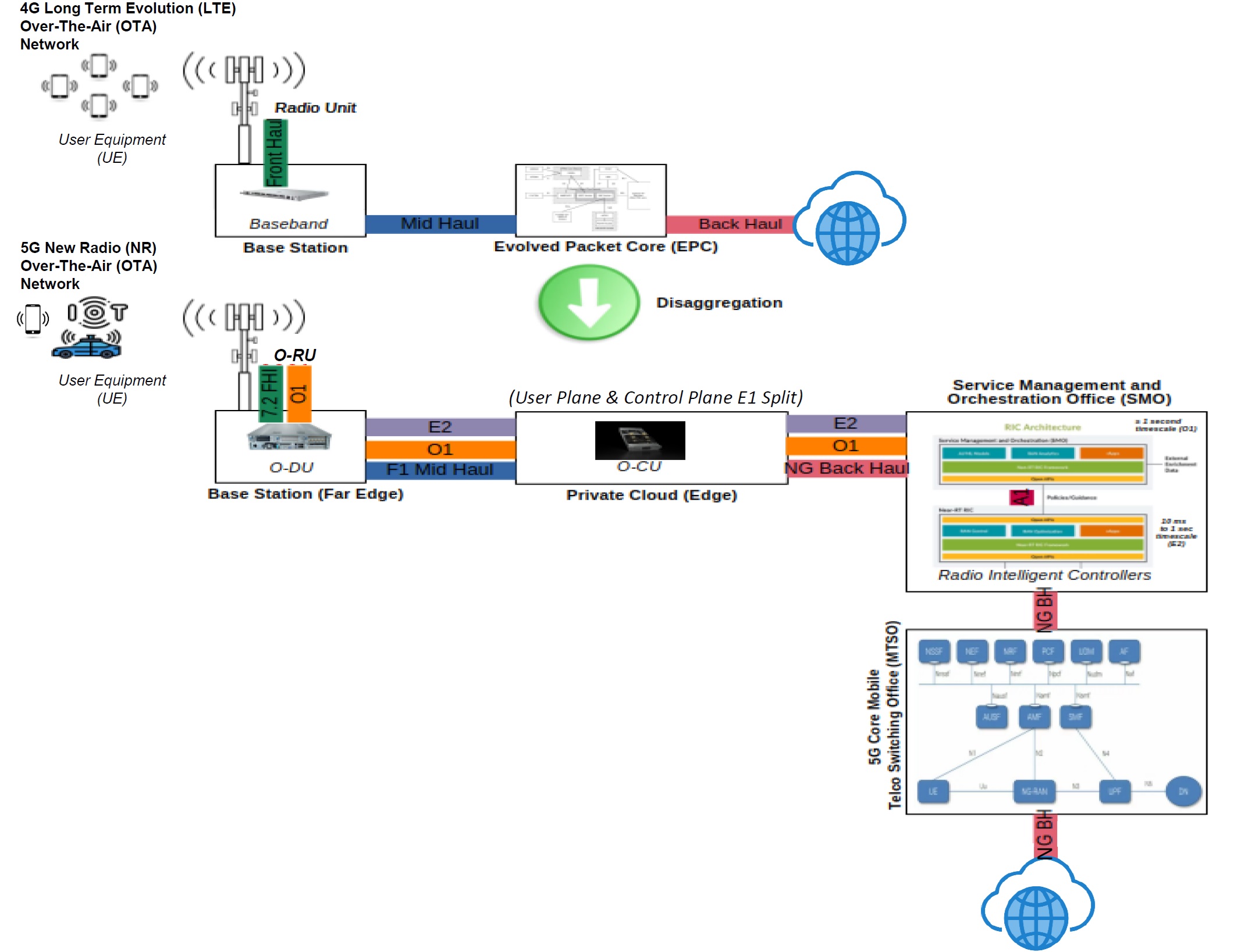} 
    \caption{Architectural evolution from monolithic RAN to Open RAN with 5G Standalone core, Service Management and Orchestration office, and disaggregated O-CU/O-DU.}
    \label{fig:evolution}
\end{figure}

\end{document}